\documentstyle[twocolumn,aps,epsf,floats]{revtex}
\begin{document}

\twocolumn[\hsize\textwidth\columnwidth\hsize\csname %
@twocolumnfalse\endcsname

\title{Effect of Superconductivity on the Incommensurate Magnetic Response 
of Cuprate Superconductors}
\author{Dirk K. Morr $^{1}$ and David Pines $^{1,2}$}
\address{ $^{1}$ University of Illinois at Urbana-Champaign,  
Loomis Laboratory of Physics, 1110 W. Green St., Urbana, IL 61801\\
$^{2}$ Center for Materials Science, LANSCE, and Center for Nonlinear Studies, Los Alamos National Laboratory, Los Alamos, NM 87545}
\date{\today}
\draft
\maketitle
\begin{abstract}
We explain the effects of  superconductivity on the incommensurate
magnetic response 
$\chi''({\bf q},\omega)$ observed in inelastic neutron scattering measurements 
on La$_{2-x}$Sr$_x$CuO$_4$. We show that a spin-fermion model
correctly describes  the frequency and momentum 
dependent changes of $\chi''$ in the superconducting phase. We find these 
changes are generic features of an incommensurate spin structure and 
the d-wave symmetry of the superconducting gap and are thus expected
for all cuprates with an incommensurate magnetic response. Our analysis
of INS experiments in La$_{2-x}$Sr$_x$CuO$_4$ up to optimal doping suggests a 
Fermi surface which is closed around $(\pi,\pi)$.

\end{abstract}
\pacs{PACS numbers: 74.20.-q, 74.25.Ha, 74.25.Jb} 
]

\narrowtext

The spin excitation spectrum in La$_{2-x}$Sr$_x$CuO$_4$  \cite{Shi89,Mas96,Lake98,Aep97,Yam98} in the normal and superconducting state has been intensively studied during the last few years by inelastic neutron scattering (INS) experiments. The normal state spectrum for  compounds with $x>0.04$ is characterized by peaks in 
$\chi''({\bf q}, \omega)$ at incommensurate wave-vectors ${\bf Q}_i=(1 \pm \delta,1) \pi$ and ${\bf Q}_i=(1,1 \pm \delta) \pi$ \cite{Shi89,Mas96,Yam98}, where $\delta$ increases with increasing doping. Recent INS experiments in the superconducting state of La$_{1.86}$Sr$_{0.14}$CuO$_4$ by Mason {\it et al.} \cite{Mas96} and La$_{1.83}$Sr$_{0.17}$CuO$_4$ by Lake {\it et al.} \cite{Lake98} show striking momentum and frequency dependent changes in $\chi''$ upon entering the superconducting state. They find for both compounds that 
at the incommensurate wave-vector ${\bf Q}_i$, 
$\chi''$ in the superconducting state is considerably decreased below its normal state value for $\omega \leq 7$ meV, while it increases for larger frequencies.
Moreover, for frequencies in the vicinity of 7 meV, the incommensurate 
peaks sharpen in the superconducting state,  while at higher frequencies 
the peak widths in the normal and superconducting state 
are approximately equal.

In this communication, we show that these experimental
results are a direct consequence of changes in the quasiparticle
damping of an incommensurate spin
structure due 
to the appearance of a d-wave gap in the superconducting state. 
We demonstrate that the available INS data provide insight into both the 
symmetry of the order parameter and Fermi
surface topology, and show 
that INS experiments on La$_{2-x}$Sr$_x$CuO$_4$ suggest a closed Fermi surface around $(\pi,\pi)$ up to optimal doping.
We find that comparable changes in $\chi''$ are to be expected 
for any cuprate superconductor with an incommensurate 
spin spectrum. Our approach should  thus apply to 
YBa$_2$Cu$_3$O$_{6+x}$, in which Tranquada 
{\it et al.}~\cite{Tra92} and Dai, Mook {\it et al.}~\cite{Dai97,Dai98} 
find an incommensurate spin structure at low frequencies. 

The starting point for our calculations is a 
spin-fermion model \cite{sfmodel} of incommensurate spin-excitations in which 
the position and width of the overdamped spin mode are determined by
its coupling to the planar quasi-particles. The description of 
the spin excitations as a relaxational mode is motivated by the fits
to NMR \cite{MMP} and INS \cite{Zha96} experiments in the normal state, 
and by the INS
experimental observation in La$_{1.86}$Sr$_{0.14}$CuO$_4$ that no 
dispersing spin mode exists for $\omega \leq 25$ meV \cite{Maspc}, 
well above the frequency range we consider here. 

In a spin-fermion model \cite{sfmodel}, the spin-wave propagator, $\chi$,  is given by 
\begin{equation}
\chi^{-1} = \chi_0^{-1} - {\rm Re}\, \Pi - i \ {\rm Im}\, \Pi \ ,
\label{Dyson}
\end{equation}
where $\chi_0$ is the bare propagator, and $\Pi$ is the irreducible particle-hole bubble. The form of the bare propagator in Eq.(\ref{Dyson}) is model dependent; however $(\chi_0^{-1} - {\rm Re}\, \Pi)$ may be taken 
from fits to NMR and INS experiments in the normal state of La$_{1.86}$Sr$_{0.14}$CuO$_4$ \cite{MMP,Zha96},
\begin{equation}
\chi_0^{-1} - {\rm Re}\, \Pi = { 1 + \xi^2 ({\bf q} - {\bf Q}_i)^2 \over 
\chi_{{\bf Q}_i} }  \ ,
\label{chifull}
\end{equation}
where $\xi$ is the magnetic correlation length and $\chi_{{\bf Q}_i}$ is 
the static staggered  susceptibility. On combining Eqs.(\ref{Dyson}) 
and (\ref{chifull}) we obtain
\begin{equation}
\chi''({\bf q}, \omega)= \chi_{{\bf Q}_i} {  \Gamma(\omega) \over
\big[1+\xi^2({\bf q}-{\bf Q}_i)^2 \big]^2 + \Gamma(\omega)^2 }  \ ,
\label{chi_full}
\end{equation}
where $\Gamma(\omega)= \chi_{{\bf Q}_i} {\rm Im }\, \Pi$.

We thus need to calculate only the imaginary part of $\Pi$ which describes the damping brought about by the decay of a spin excitation into a particle-hole pair. In the superconducting state we find to lowest order in the spin-fermion coupling $g_{eff}$ (for $\omega>0$)
\begin{eqnarray}
{\rm Im } \, \Pi({\bf q}, \omega) &=& { 3 \pi g^2_{eff} \over 8} \sum_{\bf k} 
\Big(1-n_F(E_{\bf k+q})-n_F(E_{\bf k}) \Big) \nonumber \\
& & \hspace{-1.0cm} \times 
\Bigg[ 1- { \epsilon_{\bf k+q} \epsilon_{\bf k} + \Delta_{\bf k+q} 
\Delta_{\bf k} \over E_{\bf k+q} E_{\bf k} } \Bigg]  
 \delta\Big( \omega-E_{\bf k} - E_{\bf k+q} \Big)  \nonumber \\
& & \hspace{-1.0cm} - \Big(n_F(E_{\bf k})-n_F(E_{\bf k+q}) \Big) 
\Bigg[ 1+ { \epsilon_{\bf k+q} \epsilon_{\bf k} + \Delta_{\bf k+q} 
\Delta_{\bf k} \over E_{\bf k+q} E_{\bf k} } \Bigg]  \nonumber \\
& & \hspace{-1.0cm} \times \Big\{ 
\delta \Big( \omega+E_{\bf k} - E_{\bf k+q} \Big) - 
\delta \Big( \omega-E_{\bf k} + E_{\bf k+q} \Big) \Big\} \ ,
\label{Pi}
\end{eqnarray}
where $n_F$ is the Fermi function, $E_{\bf k}= \sqrt{ \epsilon_{\bf k}^2 + |\Delta_{\bf k}|^2}$ is the fermionic dispersion in the superconducting 
state, $\Delta_{\bf k}$ is the d-wave gap
\begin{equation} 
\Delta_{\bf k}(T)=\Delta_{SC}(T) \   { \cos(k_x) - \cos(k_y) \over 2} \ ,
\end{equation} 
and $\epsilon_{\bf k}$ is the tight-binding band for a single-layer system with
\begin{eqnarray}
\epsilon_{\bf k} &=& -2t \Big( \cos(k_x) + \cos(k_y) \Big) \nonumber \\
& & \quad -4t^\prime \cos(k_x) \cos(k_y)  -\mu \ ,
\label{dispersion}
\end{eqnarray}
where $t, t^\prime$ are the hopping elements between  nearest and  next-nearest neighbors, respectively, and $\mu$ is the chemical potential.
We determine  $g_{eff}$ by requiring that for a given quasi-particle spectrum, Im$\, \Pi$ reproduces the spin-damping seen at low frequencies in NMR experiments in the normal state (where $\Delta_{SC} \equiv 0$ in Eq.(\ref{Pi}))
just above $T_c$ and we assume that  $g_{eff}$ does not change in the superconducting state. 
In what follows, we consider for definiteness La$_{1.86}$Sr$_{0.14}$CuO$_4$
\cite{Zha96}, where a choice of $t=300$ meV, $t^\prime=-0.22t$, and $\mu=-0.81t$ yields the best agreement with the experimental data. The superconducting gap, $\Delta_{SC}(T=0)\approx 10$ meV, was extracted from Raman scattering experiments by Chen {\it et al.}\cite{Chen94} and
the incommensurate wave-vector ${\bf Q}_i$ is at $\delta=0.245$
\cite{Mas96,Yam98}.

\begin{figure} [t]
\begin{center}
\leavevmode
\epsfxsize=7.5cm
\epsffile{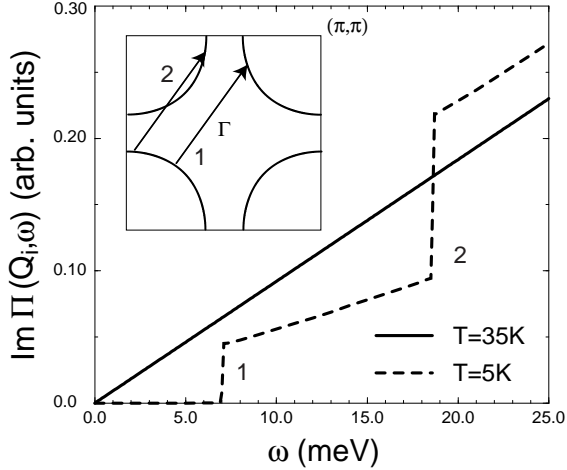}
\end{center}
\caption{ The spin-damping Im$\, \Pi$ at ${\bf Q}_i$ as a function of frequency in the normal (solid line) and superconducting state (dashed line). Inset: Fermi surface of La$_{2-x}$Sr$_x$CuO$_4$ and quasiparticle threshold transitions with wave-vector ${\bf Q}_i$.}
\label{sd_om}
\end{figure}

In Fig.~\ref{sd_om} we present our results for the spin-damping, Im$\, \Pi$, 
at the incommensurate wave-vector ${\bf Q}_i$.
Due to the incommensurate spin-structure one obtains four decay channels
for the spin excitations at ${\bf Q}_i$; the two channels within the first Brillouin zone are shown in the inset of Fig.~\ref{sd_om}. In the normal state all four channels can be excited in the low frequency limit, and one finds a linear in frequency dependence of the spin damping. In the superconducting state, the channels split into two pairs with degenerate finite threshold energies which are determined by
the momentum dependence of the order parameter and the shape of the
Fermi surface. The frequency position of the thresholds is specified by
$\omega_c^{(1,2)}=|\Delta_k|+|\Delta_{k+Q_i}|$, where $k$ and $k+Q_i$
both lie on the Fermi surface, as shown in the inset of Fig.~\ref{sd_om}. 
A comparison of the threshold energies thus provides insight into the 
symmetry of the order parameter, and in particular, is sensitive to its 
deviation from a pure d-wave symmetry.
For the band parameters we have chosen, there are two distinct threshold
energies, $\omega_c^{(1)}=0.70 \Delta_{SC}$ for quasiparticle excitations close to the nodes of the superconducting gap (excitation 1), and  $\omega_c^{(2)}=1.86 \Delta_{SC}$ for excitations close to $(0,\pi)$ and $(\pi,0)$ (excitation 2).  
At $T=0$, no quasi-particle excitations below $\omega_c^{(1)}$ can be excited, and the spin-damping vanishes. 
Both steps in Im$\, \Pi$ will, via the Kramers-Kronig relation, lead to a logarithmic divergence in Re$\, \Pi$. 
This is neglected in Eq.(\ref{chifull}), since fermion lifetime effects as found, e.g., in strong coupling scenarios \cite{Chu98}, smooth out the sharp step in Im$\, \Pi$ and thus eliminates the divergence in Re$\, \Pi$. 

\begin{figure} [t]
\begin{center}
\leavevmode
\epsfxsize=7.5cm
\epsffile{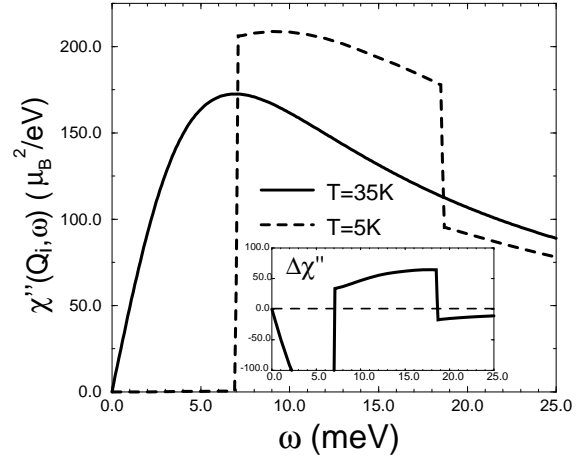}
\end{center}
\caption{ $\chi''({\bf Q}_i, \omega)$ in the normal (solid line) and superconducting state (dashed line). Inset: $\Delta \chi''(\omega)=
\chi_S''({\bf Q}_i,\omega)-\chi_N''({\bf Q}_i,\omega)$.}
\label{chi_om}
\end{figure}

In Fig.~\ref{chi_om} we present our results for the normal and superconducting state intensity $\chi''_N$  and $\chi''_S$.
Since in the normal state Im$\, \Pi \sim \omega$, we find $\chi''_N \sim \omega$ for small frequencies, followed by a maximum for $\Gamma(\omega)=1$ (which within experimental errors occurs at 
approx.~$7$ meV \cite{Mas96}), and by $\chi''_N \sim 1/\omega$ at high frequencies. On the other hand, $\chi''_S$ is characterized by jumps at $\omega_c^{(1,2)}$, which reflect the behavior of Im$\, \Pi$; an enhancement over the normal state intensity is found for the frequency range between the two jumps. This enhancement follows straightforwardly from Eq.(\ref{chi_full}) as long as the condition 
$1/\Gamma_N(\omega) \leq \Gamma_S(\omega) \leq \Gamma_N(\omega) $
(for $\chi_{{\bf Q}_i}^S/\chi_{{\bf Q}_i}^N=1$) is satisfied. From Fig.~\ref{sd_om} we find that this condition is met in a finite frequency region, leading to two crossings of $\chi_N''$ and $\chi_S''$. The location in frequency  space of the lower crossing  depends not only on $\omega_c^{(1)}$, but 
also on the ratio of  the peak strengths, 
$\chi_{{\bf Q}_i}^S/\chi_{{\bf Q}_i}^N$, or what is equivalent, to the 
ratio of the corresponding correlation lengths,  since $\chi_{{\bf Q}_i}=\alpha \xi^2$, where $\alpha$ is a temperature independent scale factor. On comparing our results with the experimental results \cite{Mas96}, which 
show a  lower crossing at $\approx 7$ meV, a resolution limited decrease of $\chi_S''$ below this crossing \cite{Lake98} and a slightly enhanced $\chi_{{\bf Q}_i}$ in the superconducting state, we conclude that a ratio $\chi_{{\bf Q}_i}^S/\chi_{{\bf Q}_i}^N \approx 1.2$ provides a reasonable fit to experiment. 
The corresponding value of $\Delta \chi''=\chi''_S-\chi''_N$ is presented in the inset of Fig.~\ref{chi_om}.

\begin{figure} [t]
\begin{center}
\leavevmode
\epsfxsize=7.5cm
\epsffile{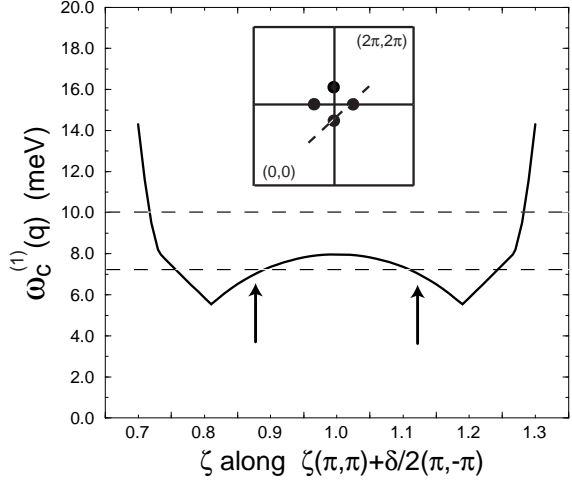}
\end{center}
\caption{ $\omega_c^{(1)}({\bf q})$ along the path shown in the inset. The arrows indicate the position of the incommensurate peaks at $\zeta=1\pm \delta/2=0.88 (1.12)$. Inset: momentum space path with ${\bf q}=\zeta (\pi,\pi)+\delta/2(\pi,-\pi)$.}
\label{om_c1}
\end{figure}

We turn next to changes in the momentum dependence of $\chi''$ at fixed frequency as one goes from the normal to the superconducting state. These changes arise from the momentum dependence of 
$\omega_c^{(1)}({\bf q})$ in the vicinity of ${\bf Q}_i$. 
In Fig.~\ref{om_c1} we plot $\omega_c^{(1)}({\bf q})$ for the momentum 
space path considered in Ref.~\cite{Mas96}, ${\bf q}=\zeta (\pi,\pi)+\delta/2(\pi,-\pi)$ (see inset). 
A momentum dependence of  $\omega_c^{(1)}({\bf q})$ which agrees
 with the experimentally observed threshold frequency for non-zero $\chi_S''$ \cite{Lake98,Maspc}, gives rise to several interesting effects. First, since
the minimum $\omega_{min}=5.5$ meV in $\omega_c^{(1)}({\bf q})$ does not occur at the incommensurate wave-vectors 
${\bf Q}_i$, a  non-zero $\chi_S''$  (at $\omega \geq \omega_{min}$) first 
appears at  wavevectors  slightly removed from ${\bf Q}_i$.
Second, as the frequency increases above $\omega_{min}$, the region in momentum space in which $\chi_S''$ is finite widens, and as a result more of the incommensurate peak, centered  at ${\bf Q}_i$, becomes "visible".
One thus obtains not only  an increase of  the peak intensity  with increasing frequency, but also a shift of the peak maximum  towards ${\bf Q}_i$, in full agreement with the experimental data \cite{Lake98,Maspc}. It follows from Fig.~\ref{om_c1} that this effect should vanish for frequencies larger than 
the local maximum of $\omega_c^{(1)}({\bf q})$ at $\zeta=1$, $\omega_{max}^{loc} = 8$ meV.

\begin{figure} [t]
\begin{center}
\leavevmode
\epsfxsize=7.5cm
\epsffile{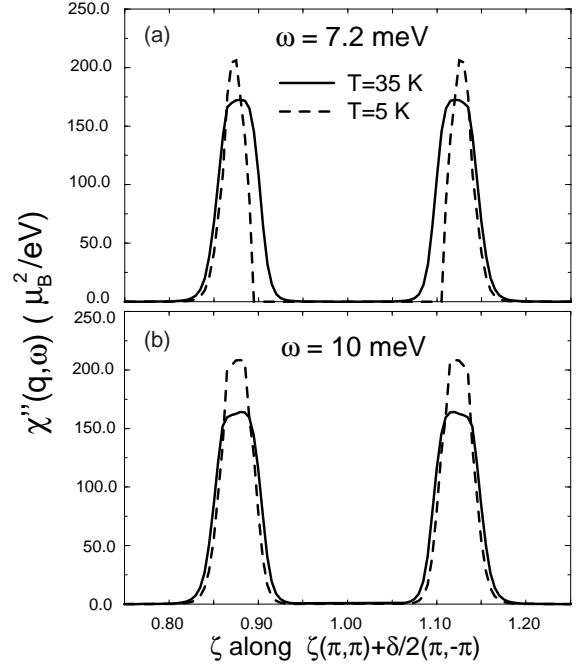}
\end{center}
\caption{  $\chi''({\bf q}, \omega)$ in the normal (solid line) and superconducting state 
(dashed line) for fixed frequency {\it (a)} $\omega=7.2$ meV  and 
{\it (b)} $\omega=10.0$ meV, along the same momentum space path as in Fig.~\protect\ref{om_c1}.}
\label{chi_q}
\end{figure}

In Fig.~\ref{chi_q} we plot the normal and superconducting $\chi''$ for $\omega=7.2$ meV  and $\omega=10.0$ meV, corresponding to the lower and upper dashed curves in Fig.~\ref{om_c1},  respectively. 
Since the frequency in Fig.~\ref{chi_q}a is between 
$\omega_c^{(1)} \equiv \omega_c^{(1)}({\bf Q}_i)$ and $\omega_{max}^{loc}$, $\chi_S''$ increases compared to the normal state result, while the width of the peak in the superconducting state is narrower than that in the normal state and the maximum of the peak intensity is shifted away from ${\bf Q}_i$. 
In contrast, for $\omega=10.0$ meV $>\omega_{max}^{loc}$ (Fig.~\ref{chi_q}b) the peak is centered around ${\bf Q}_i$
and the peak widths in the normal and superconducting state are approximately equal. All of these results, i.e., narrow peaks in the vicinity of 
$\omega_c^{(1)}({\bf Q}_i)$ as well as the increase in peak width and the shift towards ${\bf Q}_i$ with increasing frequency agree with the experimental 
findings of Ref.~\cite{Mas96,Lake98}.

An additional interesting effect appears as one moves away from ${\bf Q}_i$. Due to the symmetry of the Fermi surface, the double-degeneracy of the threshold frequencies is lifted for 
${\bf q} \not = {\bf Q}_i$. While the frequency positions of the two upper thresholds stay close, the separation between the two lower threshold energies increases rapidly with distance from ${\bf Q}_i $, e.g., for $\zeta=0.89$ the lower thresholds are located at  7.1 meV and 10.0 meV, resulting in two sharp increases of $\chi_S''(\omega)$ at these energies.  We thus predict that as one moves away from ${\bf Q}_i$, 
$\chi_S''(\omega)$ should not only decrease in intensity, but also acquire some new structure due to additional frequency thresholds. Some support for this prediction comes from our results for $\zeta=0.94$, where we obtain two sharp increases of $\chi_S''(\omega)$ at 
7.7 meV and 16 meV, with an almost constant intensity in between,  
in qualitative agreement with the experimental data \cite{Lake98}. However, due to the decrease in the signal-to-noise ratio as one moves away from 
${\bf Q}_i$ a better experimental resolution is required to allow a 
quantitative comparison with our predictions.

Within our scenario, the available INS data also provide insight into the Fermi surface topology of La$_{2-x}$Sr$_x$CuO$_4$. While we find that $\omega_c^{(2)}$ is robust against changes in the bandstructure parameters, $\omega_c^{(1)}$ depends very sensitively on the Fermi surface topology, i.e., $t^\prime$, due to the strong momentum dependence of the superconducting gap in the vicinity of the nodes. The location of $\omega_c^{(1)}$ is therefore a sensitive probe of the form of the Fermi surface and can be used to extract $t^\prime$ from the experimental data. Though the limited experimental resolution leaves some uncertainty in $\omega_c^{(1)}$, and thus in $t^\prime$,  we find that the INS data provide a lower bound for $t^\prime$. Within our scenario for the superconducting state,  particle-particle excitations in the vicinity of the nodes become impossible for $|t'|<0.2t$, hence $\chi''_S(\omega)$ would 
vanish for {\it all} frequencies below $\omega_c^{(2)}$, in contradiction to the experimental results. Moreover, assuming a weak doping dependence of $t^\prime$, this lower bound $|t'| \geq 0.2t$ yields a  Fermi surface of  La$_{2-x}$Sr$_x$CuO$_4$ which is closed around $(\pi,\pi)$ up to optimal doping. The Fermi surface thus possesses the same topology as the one in YBa$_2$Cu$_3$O$_{6+x}$; this offers a possible explanation for the occurrence of  incommensurate peaks in the spin spectrum along the same direction in momentum space in the latter 
materials \cite{Tra92,Dai97,Dai98}.
Two additional constraints for the form of the Fermi surface come from the location of the minimum in $\omega_c^{(1)}({\bf q})$ (see Fig.~\ref{om_c1}) and the appearance of additional structure in $\chi_S''(\omega)$ at ${\bf q} \not = {\bf Q}_i$,  which both strongly depend on $t^\prime$. 

It follows from the above analysis that $\omega_c^{(1)}({\bf Q}_i)$ is sensitive to the extent of incommensuration in the spin spectrum. Using the doping dependence of $\delta$ in La$_{2-x}$Sr$_x$CuO$_4$ found by Yamada 
{\it et al.} \cite{Yam98} we obtain for $x=0.1$ ($\delta=0.2$), 
$\omega_c^{(1)}({\bf Q}_i)=0.93 \Delta_{sc}$, while for $x=0.08$ ($\delta=0.16$) we find $\omega_c^{(1)}({\bf Q}_i)=1.14 \Delta_{sc}$. We thus predict that the ratio $\omega_c^{(1)}({\bf Q}_i)/\Delta_{sc}$ should increase with decreasing doping.

Though the details of the effects which we discussed are sensitive to material specific parameters, e.g., Fermi surface topology, extent of $\delta$, their experimental observation within our scenario only depends on two criteria: the existence of an incommensurate spin structure and a d-wave gap in the superconducting state; we thus predict comparable changes in $\chi_S''$ for all cuprate superconductors in which these criteria are met.

Finally, it was recently pointed out that the formation of charged stripes, which, it was suggested could be the origin of the incommensurate spin ordering,  would give rise to a substantially "smeared out" Fermi surface \cite{Sal96}. This in turn should lead to a much softer increase of 
$\chi_S''({\bf Q}_i)$ above $\omega_c^{(1)}$, in contrast to the actually very sharp and resolution limited increase observed by Lake {\it et al.} \cite{Lake98,Maspc}.

In summary, we find that the frequency and momentum 
dependent changes of $\chi_S''$ in the superconducting state are 
a direct  consequence of changes in the quasiparticle spectrum due 
to the appearance of a d-wave gap. 
We show that the available INS data also constrain the Fermi surface topology, and suggest a Fermi surface in La$_{2-x}$Sr$_x$CuO$_4$ which is closed around $(\pi,\pi)$ up to optimal doping. We predict that $\chi_S''(\omega)$ will posses additional threshold energies away from ${\bf Q}_i$, and that the ratio $\omega_c^{(1)}({\bf Q}_i)/\Delta_{sc}$ will increase with decreasing doping. Improved experimental resolution will make it possible to test our predictions. Finally, we expect to find comparable changes 
in $\chi_S''$ in all cuprate superconductors with incommensurate spin-structure. 

We would like to thank A.V. Chubukov, P. Dai, A. Millis, H. Mook,  and J. Schmalian for valuable discussions and particularly
B. Lake and T. Mason for very stimulating discussions and for providing us with their experimental data prior to publication. This work has been supported in part by the Science and Technology Center for Superconductivity through NSF-grant DMR91-20000, and by DOE at Los Alamos.

\end{document}